\documentclass[useAMS,usenatbib, usegraphicx]{mn2e}

 \usepackage{amssymb}

\title[Optical supernova remnants and H$\alpha$ star formation rates]{Optical supernova remnants
in nearby galaxies and their influence on star formation rates derived from H$\alpha$ emission}
\author[M. M. Vu\v{c}eti\'{c}, B. Arbutina \& D. Uro\v{s}evi\'{c}]{M. M. Vu\v{c}eti\'{c},$^{1}$\thanks{E-mail:
mandjelic@matf.bg.ac.rs}B. Arbutina$^{1}$ and D. Uro\v{s}evi\'{c}$^{1,2}$\\
$^{1}$Department of Astronomy, Faculty of Mathematics,
University of Belgrade, Studentski trg 16, 11000 Belgrade,
Serbia\\
$^{2}$Isaac Newton Institute of Chile, Yugoslavia Branch}
\begin{document}

\date{Accepted . Received ; in original form }

\pagerange{\pageref{firstpage}--\pageref{lastpage}} \pubyear{2014}

\maketitle

\label{firstpage}

\begin{abstract}
In this paper we present the most up-to-date list of nearby
galaxies with optically detected supernova remnants (SNRs). We
discuss the contribution of the H$\alpha$ flux from the SNRs to
the total H$\alpha$ flux and its influence on derived star
formation rate (SFR) {for 18 galaxies in our sample}. We found
that the contribution of SNRs' flux to the total H$\alpha$ flux is
$5\pm 5$\%. Due to the observational selection effects, the SNRs
contamination of SFRs derived herein represents only a lower
limit.
\end{abstract}


\section{Introduction}

Understanding and modeling the star formation rates (SFRs)
are the central goals of the theory of star formation. SFR is one
of the crucial ingredients of cosmological simulations of galaxy
formation, and these simulations demonstrate the impact of SFR
models on galaxy evolution. The SFR has been a major issue in
astrophysics since the 1970's, and during the last two decades,
the cosmic star formation history of the universe has been widely
studied in order to better constrain galaxy formation and
evolution models.\footnote{Some recent models, however, show
that galaxies self-regulate -- star formation is regulated by
stellar feedback (radiation, stellar winds and supernovae)
limiting the amount of very dense gas available for forming stars
(Hopkins, Quataert \& Murray 2011).} Large-area surveys and use of
larger telescopes have improved our knowledge of SFRs at low to
intermediate redshifts ($z<1.0$; see Hopkins \& Beacom 2006)
and beyond that (e.g. Karim et al. 2011, Sobral et al. 2013). These studies show strong decrease of SFR toward present epoch, and they suggest that the star formation activity over the last $\sim$11 Gyrs is responsible for producing $\sim$ 95 per cent of the total stellar mass density
observed locally.

Determination of SFRs in galaxies through the Hubble
sequence provides vital clues to the evolutionary histories of
galaxies. Measured SFRs are spread along six orders of
magnitude (when normalized by galaxy mass), from almost zero
in gas-poor elliptical, disk and dwarf galaxies, up to $\sim$100
M$_{\odot} \textrm{yr}^{-1}$, in optically-selected starburst
galaxies, or even more in the most luminous infrared starburst
galaxies (Kennicutt 1998).

The first quantitative SFRs were derived from evolutionary
synthesis models of galaxy colours (Searle, Sargent \& Bagnuolo
1973). Further on, more precise diagnostic tools were made using
integrated emission-line fluxes (Kennicutt 1983), near-ultraviolet
continuum fluxes (Donas \& Deharveng 1984), and infrared continuum
fluxes (Rieke \& Lebofsky 1978). Nowadays, many different
properties are used as star formation tracers, with the goal of
directly or indirectly targeting continuum or line emission that
is sensitive to the short-lived massive stars. Also, different
techniques are used depending on whether we measure SFRs for the
whole galaxy, or for regions within a galaxy (e.g. molecular
clouds). The most common approach for measuring SFRs in resolved
regions is to count individual objects (e.g. young stellar
objects) or events (e.g. supernovae) that trace the recent star
formation. Calibrations of SFR indicators have been made from the
X-rays, ultraviolet (UV), via the optical and infrared (IR), all
the way to the radio waves, using both continuum and line
emission. For the most recent review on the latest achievements in
the field see Kennicutt \& Evans (2012), as well as the previous
review by Kennicutt (1998).

In this paper, we focus on better constraining SFRs from the H$\alpha$ emission line.
The next section will introduce star formation measurements from this emission line.

\subsection{Star Formation Rates from H$\alpha$ flux}
The nebular emission lines are very effective in re-emitting
stellar luminosity, and thus provide direct measurement of
young stellar content in the galaxy. This is especially the case
for H$\alpha$ line, observationally the strongest among the
recombination lines from regions of ionized gas surrounding young
hot stars ({\hbox{H\,{\sc ii}}} regions). Only stars with masses
exceeding 10 Solar masses contribute significantly to the stellar
flux which can ionize interstellar medium (ISM). Also, these stars
have a lifetime shorter then 20 Myr, so the emission lines give us
almost  instantaneous SFRs, independent of previous star
formation histories. The conversion factor between integrated
emission-line luminosity and SFR is computed using stellar
evolutionary synthesis models. While calibrations have been
published by numerous authors, here we use the calibration from
Kennicutt, Tamblyn \& Congdon (1994), which assumes solar
abundances and the Salpeter initial mass function (IMF) (Salpeter
1955) for stellar masses in range 0.1-100M$_{\sun}$:

\begin{equation}
\\\mathrm{SFR}\ (M_{\sun}\mathrm{yr}^{-1})=7.94\times10^{-42}L_{\mathrm{H}\alpha}\ (\mathrm{erg\ s^{-1}}),
\end{equation}

\noindent where calibration coefficient yields for the Case B
recombination at $T_{\rm{e}}=10000K$. There is a significant
variation among published values for this calibration coefficient
($\sim30\%$). The differences in coefficient values originate in
the usage of different stellar evolution, atmosphere models and
IMFs.

The method for deriving SFRs from the H$\alpha$ flux is one of the
most used methods due to the advantages of optical observations
and strength of the H$\alpha$ line. The star formation in nearby
galaxies can be mapped at high resolution even with small
telescopes, and the H$\alpha$ line can be detected in the
redshifted spectra of starburst galaxies up to $z<3$
(Bechtold et al. 1997, Geach et al. 2008, Sobral et al. 2013).

The main limitations of the method are its sensitivity to
uncertainties in extinction and the IMF, and the assumption that
all of the massive star formation is traced by the ionized gas.
There is a fraction between 15\% - 50\%, of ionizing radiation
which escapes from {\hbox{H\,{\sc ii}}} regions. This can be
measured either directly (Oey \& Kennicutt 1997), or from the
observations of the diffuse H$\alpha$ emission in nearby galaxies.
In their analysis of diffuse H$\alpha$ emission Ferguson et al.
(1996) found that diffuse ionized gas is photoionized by Lyman
continuum photons which have escaped from {\hbox{H\,{\sc ii}}}
regions. Therefore, diffuse H$\alpha$ emission should be included
when this SFR measurement method is used.

The dominant source of systematic error in SFR measurements from
H$\alpha$ fluxes is extinction within the galaxy observed.
Internal extinction can be corrected by combining H$\alpha$ line
and IR continuum emission or radio data {(Kennicutt et al. 2009)},
which
 are not affected by the extinction. Mean extinction values range from A(H$\alpha$)=0.5 mag to A(H$\alpha$)=1.8 mag
(see Kennicutt (1998) and James et al. (2005) and references
therein), depending on the galaxy type and luminosity. As this
extinction correction comes with large uncertainty, and therefore
represents a large change (2 - 3 times higher) in derived fluxes,
this correction {was usually} not applied when calculating SFRs.
{With recent IR surveys (\textit{Spitzer}, \textit{Herschel},
\textit{WISE}), which give estimates on dust amount in nearby
galaxies, extinction correction is more commonly considered.}

Also, we emphasize the importance of eliminating H$\alpha$ flux
contaminants when calculating SFRs from H$\alpha$ emission.
Emission spectra from {\hbox{H\,{\sc ii}}} regions show that very
close to the H$\alpha$ line, on both sides, are \hbox{[N\,{\sc
ii}]} lines at $\lambda$654.8 nm and $\lambda$658.3 nm. Most of
the filters used to extract H$\alpha$ line also let some of the
\hbox{[N\,{\sc ii}]} emission pass through. Using different
methods, this problem can be minimized. However, it leaves some
uncertainty in derived H$\alpha$ fluxes. Using spectroscopic
observations, we can calculate the ratio between H$\alpha$ and
\hbox{[N\,{\sc ii}]} lines, and using filter profiles, we can
obtain corrections for \hbox{[N\,{\sc ii}]} contamination. Another
possibility is to use very narrow \hbox{[N\,{\sc ii}]} filters for
deriving this correction (James et al. 2005). However, because of
gradient in \hbox{[N\,{\sc ii}]} abundances with change of
galactocentric distance, one should be careful when applying any
correction. The most commonly used corrections for \hbox{[N\,{\sc
ii}]} emission for entire galaxies are those derived by Kennicutt
(1983) and Kennicutt \& Kent (1983). Using Sloan Digital Sky
Survey Data Release 4 galaxies (SDSS DR4, Adelman-McCarthy et al.
2006), a simple correction has also been derived by Villar et al.
(2008) and presented by Sobral et al. (2012) and is now widely
used.

Another origin of H$\alpha$ flux contamination are the sources
which emit H$\alpha$ radiation, but which are not {\hbox{H\,{\sc
ii}}} regions surrounding young high-mass OB stars. There is a
wide range of such objects which cause overestimate in SFRs:
another emission nebulae - planetary nebulae (PNe) and SNRs;
active galactic nuclei (AGNs); ultraluminous X-ray sources (ULXs)
and their surrounding nebulae;  {micro-quasars;} foreground stars
with H$\alpha$ emission; superbubbles. None of these types of
objects have so far been thoroughly discussed as possible sources
of systematic error in H$\alpha$ flux-based determination of SFRs.

In some papers, there are efforts to exclude emission from PNe
when making catalogues of {\hbox{H\,{\sc ii}}} regions and
estimating SFRs. Azimlu, Marciniak \& Barmby (2011) did this for
M31 galaxy, and when they removed all PNe candidates, they found
that PNe were responsible for 1\% of the total measured H$\alpha $
emission. The problem with PNe, as well as with SNRs, is that they
are hard to differentiate from {compact} {\hbox{H\,{\sc ii}}}
regions based on H$\alpha$ emission only. Additional observations
in narrow band filters can certainly distinguish between different
emission nebulae types. Also, PNe are much smaller in sizes, when
compared to the majority of {\hbox{H\,{\sc ii}}} regions. HII
regions can be compact, the same in size as PNe - smaller than a
parsec, if they are excited
 only by a single star, but {\hbox{H\,{\sc ii}}} regions are generally larger, being excited by multiple young massive stars, or even clusters. On the other hand,
SNRs can be also different in sizes, from being compact (few
parsecs), to huge (up to the 100pc; {\v{C}ajko, Crawford \&
Filipovi\'{c} (2009)}), depending on the phase of their evolution,
as well as on input energy from SN explosion and surrounding ISM
density.

Some AGNs are prominent in H$\alpha$ line from their
broad-line and narrow-line regions (it depends on the
redshift and the inclination angle). But, considering their small
size, and certainly a modest number of them which could be
detected in the projection of a single galaxy, their contribution
to the total H$\alpha$ flux of a galaxy would be negligible. On
the other hand, there are efforts to exclude AGNs from a sample of
star-forming galaxies at some higher redshift, which are studied
for SFRs (Garn et al. 2010, Villar et al. 2011).

ULX sources are compact X-ray sources that are located away from
the nucleus of their host galaxy, emitting well above the
Eddington limit of a 20M$_{\sun}$ black hole (L$_{X}$
$\sim3\times10^{39}$ erg s$^{-1}$). Recently, more and more ULX
sources have been detected with strange nebular emission
surrounding it. Some of them are IC342 X-1 source in IC342 galaxy
({Roberts et al. 2003, Abolmasov et al. 2007, Feng \& Kaaret
2008}), Holmberg IX X-1 in Holmberg IX galaxy ({Fabbiano 1988,
Gladstone, Roberts, \& Done 2009, Moon et al. 2011, Gris\'{e} et
al. 2011}) and MF 16 (nomenclature from Matonick \& Fesen 1997) in
NGC6946 {(Roberts \& Colbert 2003)}. Most likely this nebular
emission is coming from the accretion disks. Such nebulae are also
prominent in H$\alpha$ line and, as such, are potential
contaminants of H$\alpha$ flux and instantaneous SFRs. Of
course, ULX sources can be regarded as tracers of some past SFRs.
 {In NGC6946 galaxy, flux of the MF16 is responsible for 0.1\% of
the total H$\alpha$ flux, as is the case with IC342 galaxy and
IC342 X-1 source. On the other hand, as Andjeli\'{c} (2011) has
shown, H$\alpha$ derived SFRs for Holmberg IX galaxy can be
significantly changed if nebular emission from the ULX is removed
from the integrated H$\alpha$ flux of the galaxy. Emission from
HoIX X-1 is resposible for even 75\% of the
 H$\alpha$ emission in this dwarf galaxy.  We adopt H$\alpha$ flux of this object from Arbutina et al. (2009).}

{ Microquasars are X-ray binary systems, with accretion disk,
relativistic jets, and usually with strong and variable radio
emission. Also, they can have surrounding emission nebula. Large
nebula S26 (from Blair \& Long 1997) in the nearby galaxy NGC7793
is a jet-inflated bubble around a powerful microquasar (Pakull,
Soria \& Motch 2010). H$\alpha$ emission from this source is not
tracer of a star formation, and it is responsible for 1.15\% of
the galaxy's total H$\alpha$ emission.}

Foreground emission-line stars, such as O and B supergiants and
Wolf-Rayet stars exhibit strong optical emission lines, primarily
hydrogen Balmer lines. In most of these systems, emission
originates from strong stellar winds. While these stars have
strong emission, they also have very bright continuum, and
therefore they should be easily distinguished from {\hbox{H\,{\sc
ii}}} regions. In that sense, foreground emission-line stars could
be easily removed from the total H$\alpha$ flux and should not
represent a source of error in H$\alpha$ derived SFRs.

Superbubbles are regions of bright emission which frequently
surround OB association. This kind of nebulae is powered by a
combination of stellar winds, UV stellar radiation and occasional
SN explosions. On the other hand, wind-blown nebulae are related
to the {\hbox{H\,{\sc ii}}} regions, but they are ionized due to
shock fronts caused by strong stellar winds. Superbubbles are
characterized by low velocity shock fronts {($<$ 100 km
s$^{-1}$)}, while classical SNRs have shock fronts expanding at
100 - 1000 km s$^{-1}$. With modern observational equipment,
superbubbles can be distinguished from the SNRs, mostly upon
detection of OB associations inside. {In M31 SNR survey, Lee \&
Lee (2014) separated 44 superbubbles from previous SNR candidates,
which represent 1.3\% of total H$\alpha$ emission of this galaxy,
while in M101 galaxy 10 superbubbles represent 0.3\% (Franchetti
et al. 2012).}

%Superbubbles and diffuse ionized gas - Long at al. 2010 strana 20; Blair \& Long (1997) strana 16; Pannuti et al. 2002 strana 3. Franchetti et al. 2012.

The main aim of this paper is to estimate the influence of SNRs' emission on the H$\alpha$ derived SFRs. In order to do so, we assemble current sample
of optically detected SNRs in nearby galaxies.

{In the next section we discuss optical detection of SNRs. Section
3 gives details on individual galaxies in our sample and SNR
detections in those galaxies,
 while in section 4 we discuss influence of SNRs' emission on the H$\alpha$ derived SFRs. In section 5 we present our conclusions.
}

\section{Optical detection of SNRs}

Optical extragalactic searches for SNRs were pioneered by
Mathewson \& Clarke (1973a). They used the fact that the optical
spectra of SNRs have elevated \hbox{[S\,{\sc ii}]} $\lambda$671.7
nm, $\lambda$673.1 nm to H$\alpha$ $\lambda$656.3 nm emission-line
ratios, as compared to the spectra of normal {\hbox{H\,{\sc ii}}}
regions. This emission ratio has been used to differentiate
between shockheated SNRs (ratios $>$0.4, but often considerably
higher) and photoionized nebulae ($<$0.4, but typically $<$0.2)
(Blair \& Long 2004). This is justified through different ways of
excitation. For SNRs we have collisional excitation inducted by
shocks, rather than by photoionization, which is the case with
{\hbox{H\,{\sc ii}}} regions. In typical {\hbox{H\,{\sc ii}}}
regions, sulfur exists mainly in the form of S++, yielding low
\hbox{[S\,{\sc ii}]} to H$\alpha$ emission ratios. In the case of
SNRs, after the shock wave from an SNR propagates through the
surrounding medium and as the material cools sufficiently (when an
SNR is in the radiative phase), a variety of ionization states are
present, including S+. This is the phase when the SNRs are most
prominent in optical wavelengths and when we expect increased
\hbox{[S\,{\sc ii}]}/H$\alpha$ ratios to be observed in SNRs.

The "classical" definition for detection of an SNR at optical wavelengths is \hbox{[S\,{\sc ii}]}/H$\alpha$$>$0.4. Since the \hbox{[N\,{\sc ii}]}
lines mentioned above may be as strong as $\sim$50\% of H$\alpha$ line, some authors have relaxed the above criterion for identification
of an SNR to the condition \hbox{[S\,{\sc ii}]}/H$\alpha$$>$0.3. (Dopita et al. 2010a and Leonidaki, Boumis \& Zezas 2013).

\subsection{Optical observations of SNRs}

The optical search for SNRs is usually performed using the narrow
band H$\alpha$ and \hbox{[S\,{\sc ii}]} filters and some continuum
filter used to remove contribution from the continuum radiation.
Modern instruments, both telescopes and CCDs, have much improved
possibilities for this kind of observations in recent 15 - 20
years. That is why we have seen a significant increase in the
number of optically detected SNRs in this period. In recent years
a significant step forward has been made in making more complete
SNR samples for some galaxies, such as {M31}, M33 and M83.
Earlier, most optical detections of SNRs were performed using the
photographic plates, which had much lower sensitivity and more
difficulties in processing. The largest contribution to the number
of optically detected SNRs in this early period can be attributed
to D'Odorico, Dopita \& Benvenuti (1980) and Mathewson et al.
(1983, 1984, 1985). Mathewson and his group gave major
contribution to the detection in Magelanic Clouds (MCs), while
D'Odorico et al. (1980) observed eight galaxies - NGC6822, IC1613,
M31, M33, NGC253, NGC300, NGC2403, IC342. Both of these groups
provided only detections of SNRs, without any flux measurements.

Next major contribution to this field was also by two groups of
authors - Blair \& Long (1997) and Matonick \& Fesen (1997). The
first group observed two galaxies in Sculptor group - NGC300 and
NGC7793, while Matonick \& Fesen (1997) observed five nearby
spiral galaxies - NGC5204, NGC5585, NGC6946, M81, and M101. Also,
the same year Matonick et al. (1997) published observations of
spiral galaxy NGC2403.

Matonick \& Fesen (1997) used 1.3m McGraw-Hill Telescope at the
Michigan-Dartmouth-MIT Observatory. They took sets of three images
through each filter, with exposure times ranging from 300 to 1200
s. They also made spectroscopic observations, which they used both
to confirm detections and to calibrate H$\alpha$ flux for the
\hbox{[N\,{\sc ii}]} contamination.

Blair \& Long (1997) made observations using 2.5m du Pont Telescope at Las Campanas Observatory. They managed 1500s exposure times
in H$\alpha$ filter (but it should be noted that their filters were with little lower transmittance of 50 - 60\%), while exposure
times were at least twice as long in \hbox{[S\,{\sc ii}]} filter as in H$\alpha$, to permit comparable signal-to-noise ratios in objects having
\hbox{[S\,{\sc ii}]}/H$\alpha$ ratios of $\sim0.5$. They applied a flat correction of 25\% to remove \hbox{[N\,{\sc ii}]} contamination from the H$\alpha$ images.

Lately, Leonidaki et al. (2013) gave a major contribution to the
total number of optically detected SNRs in nearby galaxies. In six
galaxies - NGC2403, NGC3077, NGC4214, NGC4395, NGC4449 and NGC5204
they detected more than 400 SNRs. They made multiwavelength
analysis of SNRs in these six galaxies, using both their optical
and X-ray detections published previously by Leonidaki, Zezas \&
Boumis (2010). Leonidaki et al. (2013) obtained optical images
with the 1.3m Ritchey-Chr\'etien telescope at the Skinakas
Observatory. They used 3600s exposure time in H$\alpha$ filter and
7200s in \hbox{[S\,{\sc ii}]} filter. In order to estimate the
H$\alpha$ flux of their objects, they made correction for the
\hbox{[N\,{\sc ii}]} contamination using the \hbox{[N\,{\sc
ii}]}($\lambda\lambda$6548, 6584)/H$\alpha$ ratios from integrated
spectroscopy of the galaxies from the work of Kennicutt et al.
(2008).

Also, over the last 10 years, there are several studies which used
very large telescopes in SNR surveys. This resulted in making
several excellent SNR samples in nearby galaxies such as {M31},
M33 and M83. {Lee \& Lee (2014) used data provided by the Local
Group Survey (LGS; Massey et al. 2006) obtained by 4 m Mayall
telescope at Kitt Peak National Observatory (KPNO)}. For detection
of SNRs in M33, Long et al. (2010) used the same telescope,
complemented with very deep exposures using 0.9 m Burrell Schmidt
telescope. M83 galaxy was observed by using the same 4 m telescope
by Blair \& Long (2004), but also with \textit{HST} (Dopita et al.
2010, {Blair et al. 2014}) and 6.5 m Magellan telescope (Blair,
Winkler \& Long 2012). This was enough to make M83 the galaxy with
the best sample of optically selected SNRs.

\subsection{Galaxies with optically identified SNRs}
In order to estimate contribution from SNRs to the total H$\alpha$
emission used to determine SFRs in a galaxy, we have searched
literature for all galaxies which have optically identified SNRs.
In total, there are {25} galaxies with optically detected SNRs
(excluding the Milky Way). In Table 1, we list all the 25 galaxies
which have been searched for optical SNRs, with basic data for
each galaxy taken from the Nasa Extragalactic Database,
NED\footnote{http://ned.ipac.caltech.edu/}.

\begin{table*}
 \caption{Data for galaxies which have been observed for optical SNRs.}
 \label{symbols}
 \begin{tabular}{@{\extracolsep{0mm}}lcccccccccccc}
  \hline
  Galaxy & R.A. & Decl. & Distance & Distance & Major & Minor & Galactic & Incl.$^*$ & Galaxy & {$B$} & $A_{{B}}$\\
  name & (J2000) & (J2000) & & reference & axis & axis & latitude &  &  & magnitude &  &  \\
  & (h:m:s) & (d:m:s) & (Mpc) & & (\arcmin) & (\arcmin) & (\degr) & (\degr) &  & (mag) & (mag) \\
\hline
LMC &   05:23:34.5  &   -69:45:22   &   0.05    &   1   &   645 &   550 &   -32.9   &   35  &   SB(s)m  &   0.9 &   0.272\\
SMC &   00:52:44.8  &   -72:49:43   &   0.06    &   2   &   320 &   185 &   -44.3   &   58  &   SB(s)m pec  &   2.7 &   0.134\\
NGC6822 &   19:44:57.7  &   -14:48:12   &   0.50    &   1   &   15.5    &   13.5    &   -18.4   &   33  &   IB(s)m  &   9.31    &   0.855\\
NGC185  &   00:38:58.0  &   48:20:15    &   0.62    &   3   &   11.7    &   10.0    &   -14.5   &       &   E3 pec  &   10.1    &   0.667\\
IC1613  &   01:04:47.8  &   02:07:04    &   0.65    &   4   &   16.2    &   14.5    &   -60.6   &   29  &   IB(s)m  &   9.88    &   0.090   \\
IC342   &   03:46:48.5  &   68:05:47    &   3.30    &   5   &   21.4    &   20.9    &   10.6    &   25  &   SAB(rs)cd   &   9.1 &   2.024\\
NGC253  &   00:47:33.1  &   -25:17:18   &   3.94    &   6   &   27.5    &   6.8 &   -87.9   &   85  &   SAB(s)c &   8.04    &   0.068   \\
\hline
M31 &   00:42:44.3  &   41:16:09    &   0.79    &   4   &   190 &   60  &   -21.6   &   78  &   SA(s)b  &   4.36    &   0.225   \\
M33 &   01:33:50.9  &   30:39:37    &   0.84    &   4   &   70.8    &   41.7    &   -31.3   &   54  &   SA(s)cd &   6.27    &   0.15\\
NGC300  &   00:54:53.5  &   -37:41:04   &   2.0 &   4   &   21.9    &   15.5    &   -79.4   &   45  &   SA(s)d  &   8.95    &   0.046   \\
NGC4214 &   12:15:39.2  &   36:19:37    &   2.92    &   7   &   8.5 &   6.6 &   78.1    &   39  &   IAB(s)m &   10.24   &   0.079   \\
NGC2403 &   07:36:51.4  &   65:36:09    &   3.22    &   4   &   21.9    &   12.3    &   29.2    &   57  &   SAB(s)cd    &   8.93    &   0.145   \\
M82 &   09:55:52.7  &   69:40:46    &   3.53    &   8   &   11.2    &   4.3 &   40.6    &   69  &   I0 edge-on  &   9.3 &   0.567       \\
M81 &   09:55:33.2  &   69:03:55    &   3.63    &   4   &   26.9    &   14.1    &   40.9    &   62  &   SA(s)ab &   7.89    &   0.291   \\
NGC3077 &   10:03:19.1  &   68:44:02    &   3.82    &   9   &   5.4 &   4.5 &   41.6    &   38  &   I0 pec  &   10.61   &   0.243   \\
NGC7793 &   23:57:49.8  &   -32:35:28   &   3.91    &   6   &   9.3 &   6.3 &   -77.2   &   48  &   SA(s)d  &   9.98    &   0.053   \\
NGC4449 &   12:28:11.1  &   44:05:37    &   4.21    &   9   &   6.2 &   4.4 &   72.4    &   45  &   IBm &   9.99    &   0.053   \\
M83 &   13:37:00.9  &   -29:51:56 & 4.47    &   10  &   12.9    &   11.5    &   31.9    &   28  &   SAB(s)c &   8.2 &   0.241   \\
NGC4395 &   12:25:48.8  &   33:32:49    &   4.61    &   9   &   13.2    &   11.0    &   81.5    &   34  &   SA(s)m? &   10.64   &   0.063   \\
NGC5204 &   13:29:36.5  &   58:25:07    &   4.65    &   9   &   5.0 &   3.0 &   58.0    &   54  &   SA(s)m  &   11.73   &   0.045   \\
NGC5585 &   14:19:48.2  &   56:43:45    &   5.7 &   11  &   5.8 &   3.7 &   56.6    &   51  &   SAB(s)d &   11.2    &   0.057   \\
NGC6946 &   20:34:52.3  &   60:09:14    &   5.9 &   12  &   11.5    &   9.8 &   11.7    &   32  &   SAB(rs)cd   &   9.61    &   1.241   \\
M101    &   14:03:12.5  &   54:20:56    &   6.7 &   4   &   28.8    &   26.9    &   59.8    &   22  &   SAB(rs)cd   &   8.31    &   0.031   \\
M74 &   01:36:41.7  &   15:47:01    &   7.3 &   13  &   10.5    &   9.5 &   -45.7   &   20  &   SA(s)c  &   9.95    &   0.254   \\
NGC2903 &   09:32:10.1  &   21:30:03    &   8.9 &   14  &   12.6    &   6.0 &   44.5    &   64  &   SAB(rs)bc   &   9.68    &   0.113   \\
\hline
 \end{tabular}
\begin{flushleft}
$^{*}$ From Tully (1988).\\
\textsc{Distance references:} (1) van den Bergh 2000; (2) Ferrarese et al. 2000; (3) Conn et al. 2012; (4) Freedman et al. 2001; (5) Saha, Claver \& Hoessel 2002;
(6) Karachentsev et al. 2003a; (7) Tully et al. 2006; (8) Sakai \& Madore 1999; (9) Karachentsev et al. 2003b; (10) Thim et al. 2003;
(11) Karachentsev, Kopylov \& Kopylova 1994; (12) Karachentsev et al. 2004; (13) Sharina, Karachentsev \& Tikhonov 1996; (14) Drozdovsky \& Karachentsev 2000.
\end{flushleft}

 \end{table*}

 {We note that in Table 1 there are galaxies, such as NGC253 and IC342, which do not have published flux estimates for their SNRs, or NGC185 galaxy
  for which we did not find any published H$\alpha$ flux. Also, there are galaxies (such are Large Magellanic Cloud - LMC, Small Magellanic Cloud - SMC, NGC6822
   and IC1613) which have not been surveyed for optical SNRs, but have individual detections of SNRs presented in the literature. None of these seven
   galaxies listed in the first part of Table 1 will not be included in our discussion on SNRs contribution to the H$\alpha$ derived SFRs. Anyway, we want to
    mention a few details on optical detection of SNRs in each of these galaxies, as a guide for future work in this field.}

{LMC and SMC are the closest neighbors of the Milky Way. Due to
their proximity, studies of SNRs in these galaxies began as early
as in the 1960s. At the distance of MCs (50 - 60 kpc), SNRs are
well resolved, which gives an opportunity for a more in-depth
analysis. This is why an increasing number of recent papers on
SNRs in MCs are discussing individual objects (e.g. Bozzetto et
al. 2012a; Bozzetto et al. 2012b) but through multifrequency
analysis (e.g. Bozzetto et al. 2012c; de Horta et al. 2012; Haberl
et al. 2012a; Maggi et al. 2012; Bozzetto et al. 2013; Bozzetto et
al. 2014).}

{First detections of SNRs in MCs were done by using a combination
of radio and optical techniques (Mathewson \& Clarke 1972;
 Mathewson \& Clarke 1973a; Mathewson \& Clarke 1973b; Mathewson et al. 1983, 1984, 1985).}

{Filipovi\'{c} et al. (1998) listed all the discrete radio sources
in the Parkes survey of the MCs, and found 32 SNRs and 12 SNR
candidates in the LMC, and 12 SNRs in the SMC. Filipovi\'{c} et
al. (2005) presented 16 Australian Telescope Compact Array (ATCA)
radio SNRs in SMC. Twelve of them have
 optical long-slit spectra and estimated line ratios presented by Payne et al. (2007). According to Payne, White \& Filipovi\'{c} (2008), a revision of
 the Parkes survey complemented with ATCA data yields 52 confirmed SNRs and 20 candidates in the LMC.}

{In the X-rays, Williams et al. (1999) published an atlas of
\textit{ROSAT} sources, which contained 31 LMC SNRs, while Haberl
et al. (2012b)
 were the first to cover the full extent of the SMC in X-rays.}

{Badenes, Maoz \& Draine (2010) discussed size distribution of
SNRs in MCs. In their paper they tried to collect full sample of
SNRs in MCs. They merged two online catalogues to finally obtain
54 confirmed SNRs in LMC and 23 in SMC. The first catalogue they
used is an optical catalogue of MC SNRs being assembled as part of
the Magellanic Clouds Emission Line Survey (MCELS, Smith, Leiton
\& Pizarro 2000); the most recent on-line version of this
catalogue\footnote{http://www.ctio.noao.edu/mcels/snrs/framesnrs.html}
lists 31 SNRs in the LMC and 11 in the SMC. The second catalogue
is presented by the Magellanic Clouds Supernova Remnant Database
(Murphy Williams et al. 2010)\footnote{http://www.mcsnr.org/}; the
current version of their catalogue contains 47 confirmed objects
in the LMC and 16 in the SMC. Unfortunately, none of these
catalogues contain data on optical fluxes. }

{NGC6822 is one of the nearest dwarf irregular galaxies in the
Local Group. Despite its convenient distance, only one SNR has
been detected so far. This SNR - Ho12 was first observed by Hodge
(1977), and later by D'Odorico et al. (1980) and D'Odorico \&
Dopita (1983). Kong, Sjouwerman \& Williams (2004) published
multiwavelength analysis of this SNR.}

{NGC185 is a dwarf elliptical companion of the Andromeda galaxy,
which has undergone recent star formation activity. Gallagher,
Hunter \& Mould (1984) obtained spectra of NGC185, and found
strong \hbox{[S\,{\sc ii}]} emission, which originated from an
SNR. Later, Mart\'{i}nez-Delgado, Aparicio \& Gallart (1999)
obtained new H$\alpha$ observations of the central region of
NGC185, where they also detected this SNR. Finally, Gon\c{c}alves
et al. (2012) confirmed this object as SNR, using a deep
spectroscopy.}

{IC1613 is a low mass irregular galaxy in the Local Group. The
only known SNR in this galaxy is the bright nebula S8 (Sandage
1971). This SNR was observed by numerous authors (D'Odorico et al.
1980, D'Odorico \& Dopita 1983, Peimbert, Bohigas, \&
Torres-Peimbert 1988) but complete multiwavelength analysis was
presented by Lozinskaya et al. (1998).}

{NGC253 is one of the nearest starburst galaxies. SNRs detection
in this galaxy is difficult because of the edge-on orientation.
Most of SNR surveys in this galaxy have been performed in radio
domain (Ulvestad 2000; Lenc \& Tingay 2006). In optical, so far
there has been only detection of two SNR candidates made by
D'Odorico et al. (1980).}

{IC342 is an almost face-on spiral galaxy of large angular extent.
It is heavily obscured by Galactic disk, and this is why it was
often avoided for optical observations. D'Odorico et al. (1980)
were the first who searched for IC342 SNRs at optical wavelengths.
Their paper reported detection of 4 SNRs, but they observed only
central part of the galaxy. Vu\v{c}eti\'{c} et al. (2013)
performed search for SNRs and {\hbox{H\,{\sc ii}}} regions in
South-western part of this galaxy through H$\alpha$ and
\hbox{[S\,{\sc ii}]} filters. Until now, Vu\v{c}eti\'{c} et al.
(2013) have published only detections of {\hbox{H\,{\sc ii}}}
regions, while results for SNRs are in preparation.}

\subsection{Sample selection}
{From the 25 galaxies which have optical SNRs, we have selected 18
which have been fully or partially surveyed for SNRs in optical
wavelengths and which have published flux estimates for SNRs.
Those galaxies are listed in the second part of Table 1.}

When we look at the morphologies of the galaxies in the sample,
{14} of them are spirals and {4} are irregular galaxies. Knowing
that the spiral and irregular galaxies are the places where the
most star-births are taking place, it is hardly surprising that
such galaxies have been chosen to be targets for optical searches
for SNRs. Most of the galaxies are with lower inclinations, in
order to reduce internal extinction, with the exception of few
bright and well-known galaxies, such as M31 and M82. In the same
manner, almost all galaxies (except NGC6946) avoid Galactic plane,
having Galactic latitudes $|b|>20\degr$, in order to reduce
Galactic extinction. {Also, all of galaxies in our sample have
blue apparent magnitude $B<12$.} Considering distances of galaxies
in our sample, half of them are in the range between 3 -- 5 Mpc.
The most distant galaxy with optically detected SNRs is NGC2903,
at the distance of nearly 9 Mpc.

 {We queried Nearby Galaxies Catalog (Tully 1988) to see how many galaxies
 are there which meet the criteria of our sample -- to be spiral or irregular galaxy; to have $d<9$ Mpc, $i<70\degr$, $|b|>20\degr$ and $B<12$. In total,
 there are 50 galaxies that meet these criteria, while there are 17 within the distance range of 3 -- 5 Mpc. Hence, our sample represents roughly
 35\% of galaxies which could be observed for SNRs. { The criteria that are met in our sample are simply a consequence of a bias -- observers tend to choose target objects
 which are closer, brighter and with less extinction.} If we extended conditions to cover all inclinations and to reach blue apparent magnitude of
 $B=15$, 154 galaxies would be available.}

In the sample there are galaxies (like NGC2403, M83 and M101) which have multiple detections of the same SNR by different studies. In such cases, we have taken
 flux data from the most reliable and most recent observations. Also, in the cases (such are M83 and M101) where there are observations of different parts
of the same galaxy, we naturally present SNRs detected in all of the observed parts.

{In order to compare fluxes of SNRs with H$\alpha$ fluxes of
galaxies, we correct fluxes of SNRs for Galactic extinction and
\hbox{[N\,{\sc ii}]} contamination in the same manner as for
fluxes of galaxies (more details in section 4). Such corrected
H$\alpha$ fluxes are presented in Section 4, while in the next
section we give SNRs' fluxes for each galaxy as they were
originally published.}

\section{Nearby galaxies with optical SNRs}

{In this section, we review individual galaxies in the sample and
their properties. For each galaxy we summarize the history of SNRs
detection, especially in optical range, and we specify characteristics of some unusual sources.
We have also collected the H$\alpha$  fluxes of all optically detected SNRs from
 each galaxy in the sample (which are, together with coordinates, 
diameters and \hbox{[S\,{\sc ii}]}/H$\alpha$ ratios, given in Appendix).
}

\subsection{M31}

M31 - Andromeda galaxy is the largest galaxy in the Local Group
and the nearest spiral galaxy to the Milky Way. In many
properties, this galaxy is very similar to our own, and that is
one of the reasons why it is so interesting to us. It should be
mentioned that this galaxy has small inclination to the line of
sight ($\sim13^\circ$), so it is a limiting factor for the number
of detectable SNRs. The first discussion on optical SNRs in M31
were made by Kumar (1976). DDO80 detected 19 optical SNRs, while
Blair, Kirshner, \& Chevalier (1981) reported 18, with 11 of them
having measured line fluxes. Braun \& Walterbos (1993) observed a
large fraction of the spiral arms in the Northeast half of M31 and
they detected 52 SNR candidates, both in optical and radio
wavelengths. Two years later, Magnier et al. (1995) published 178
SNR candidates, from about one square degree in galactic disk, but
they did not measure objects' fluxes. At radio wavelengths, Dickel
et al. (1982) and Dickel \& D'Odorico (1984) searched for radio
counterparts to the optical identifications, and they were
successful in total 8 detections. Kong et al. (2003) and Williams
et al. (2004) both detected two SNRs, through all three ranges of
interest - radio, optical and X-rays, but with roughly calibrated
optical data. In X-rays, Supper et al. (2001) detected 16 SNRs
using \textit{ROSAT}, while 21 were identified with
\textit{XMM-Newton} (Pietsch, Freyberg \& Haberl 2005). Lately,
Stiele et al. (2011) published catalog of \textit{XMM-Newton}
sources from entire M31, out of which 25 sources were identified
as already known SNRs, while additional 31 were classified as SNR
candidates. Sasaki et al. (2012) searched the Local Group Galaxy
Survey (LGGS\footnote{Data from the LGGS are available at
http://www.lowell.edu/users/massey/lgsurvey/.}, Massey et al.
2006) and \textit{HST} archival images for optical counterparts to
the SNRs and SNR candidates from Stiele et al. (2011). They found
that 21 X-ray SNRs have an optical counterpart, as well as 20
X-ray SNR candidates. Also, they suggested additional 5 X-ray
sources from Stiele et al. (2011) to be SNRs, based on their
optical properties.

{Just recently, Lee \& Lee (2014) presented a survey of optically
emitting SNRs in the entire disk of M31 based on H$\alpha$ and
\hbox{[S\,{\sc ii}]} images in LGS. They published a catalogue of
156 SNR candidates in M31. In their survey, they rejected all
objects with \hbox{[S\,{\sc ii}]}/H$\alpha$$>0.4$, but which also
have a number of OB stars in their projection, or have sizes of
$D>100$ pc. These objects could be {\hbox{H\,{\sc ii}}} regions or
superbubbles. This is how they made 154 previously known SNRs to
be non-SNR objects. Out of these 154 non-SNRs, 44 are catalogued
as superbubbles. Out of 156 SNRs in the catalogue, 80 objects
previously detected by Braun \& Walterbos (1993), Magnier et al.
(1995) and  Sasaki et al. (2012), are confirmed to be SNRs.}

Total H$\alpha$ flux of {156} SNRs detected in M31 by Lee \& Lee
(2014) is ${4.81}\times10^{-12}$ erg cm$^{-2}$ s$^{-1}$.

\subsection{M33}

M33 is the nearest late-type spiral galaxy in the Local group. Its
distance of 840 kpc, moderate inclination ($i=54^\circ$) and a low
Galactic absorption (A$_{{B}}$=0.15 mag, NED) make this galaxy a
good target for SNRs search. This could be the reason why M33 is
among galaxies with the largest sample of detected SNRs. It is
surely a galaxy with the largest sample of SNRs detected both in
optical and X-rays. Long et al. (2010) made an enormous
contribution to this topic by reporting and giving detailed
description of 137 SNRs in X-ray, optical and radio wavelengths.
They improved the sample of optical SNRs previously made by Gordon
et al. (1998), which consisted of 98 SNRs, detected with Kitt Peak
4 m telescope. Long et al. (2010) primarily presented results of
the ChASeM33 survey (\textit{Chandra} survey, Plucinsky et al.
2008), but they also made a valuable supporting multiwavelength
analysis. They used LGGS (Massey et al. 2006) as well as their own
optical observations made with 0.6 m Burrell Schmidt telescope,
but with very long exposures (5100 s). They made a sample of 137
SNRs, all visible in optical range, out of which 131 SNRs were
detected also in X-rays. They made new radio observations with
VLA, but in the high-resolution mode in order to complement
previous VLA observations made by Gordon et al. (1999). Gordon et
al. (1999) detected 53 SNRs, all of which were previously
suggested as SNRs based on optical imagery. Long et al. (2010)
were successful in detecting small-diameter, young remnants and
pulsar-wind nebulae (PWNe) candidates and in some cases they
resolved confusion of {\hbox{H\,{\sc ii}}} regions with SNRs.

Based on their optical observations, we found that total H$\alpha$
flux from 137 SNRs detected in M33 is $5.8\times10^{-12}$ erg
cm$^{-2}$ s$^{-1}$.

\subsection{NGC300}

NGC300 is a spiral galaxy, member of the Sculptor group of
galaxies. NGC300 is very prominent in optical and H$\alpha$ due to
many large {\hbox{H\,{\sc ii}}} regions which are evidence of
ongoing star formation (see Butler, Mart\'{i}nez-Delgado \&
Brandner 2004). First optical detection of SNR candidates was done
by D'Odorico et al. (1980). Later studies (Blair \& Long (1997),
Pannuti et al. 2000, Payne et al. 2004,  Millar et al. 2011) were
all, with the exception of Blair \& Long (1997), multiwavelength
analyses. Blair \& Long (1997) performed optical narrowband
imaging and spectroscopy and they proposed 28 SNR candidates.
Pannuti et al. (2000) presented their VLA observations, in
combination with their own optical observations, archival
\textit{ROSAT} X-ray data and optical data from Blair \& Long
(1997). They reported 17 radio SNR candidates, of which 3 were
known from the optical observations. Payne et al. (2004) gave
detailed cross-matching of objects previously detected in NGC300
and detections which they made using ATCA and \textit{XMM-Newton}
observations. Finally, Millar et al. (2011) performed spectroscopy
for 51 sources chosen from the Blair \& Long (1997) and Payne et
al. (2004), as well as analysis of archival images from
\textit{Chandra} X-ray observatory. Based on their spectroscopical
analysis, Millar et al. (2011) confirmed 22 out of 28 SNR
candidates from Blair \& Long (1997). In addition, three of them
are also visible in radio, and two were detected also in X domain.

Total integrated H$\alpha$ line flux density from those 22 confirmed SNRs from Millar et al. (2011) is
$2.3\times10^{-13}$ erg cm$^{-2}$ s$^{-1}$.

\subsection{NGC4214}

NGC4214 is a nearby irregular dwarf galaxy currently experiencing
a high level of massive star formation. Combining optical,
near-IR, and UV data, Huchra et al. (1983) concluded that it went
through a burst of star formation a few times $10^{7}$ yr ago.
This galaxy is one of the several galaxies in which surveys for
SNRs have been carried out through X-ray, radio and optical
wavelengths. Initially, Vukoti\'{c} et al. (2005) classified one
radio source as SNR from VLA archival observations of NGC4214.
Afterwards, Chomiuk and Wilcots (2009) found six more radio SNR
candidates, and three more objects denoted as SNR/{\hbox{H\,{\sc
ii}}}, also from the VLA observations. Then Leonidaki et al.
(2010) performed search for X-ray SNR candidates through archival
\textit{Chandra} images, and suggested 11 sources as SNRs or SNR
candidates. The same year, Dopita et al. (2010b) detected all
radio SNRs from Chomiuk and Wilcots (2009) using \textit{HST}
telescope. Finally, Leonidaki et al. (2013) detected 92 optical
SNRs or SNR candidates, using \hbox{[S\,{\sc ii}]}$/$H$\alpha$
ratio criterion.

These 92 optical SNR candidates in NGC4214 have a total H$\alpha$
flux of $1.73\times10^{-12}$ erg cm$^{-2}$ s$^{-1}$.

\subsection{NGC2403}

NGC2403 is a spiral galaxy at the distance of 3.22 Mpc and with the inclination angle of around 60${^\circ}$.
This galaxy is the second brightest galaxy in the M81 Group (Karachentsev et al. 2002). It was one of the targets of surveys by Leonidaki et al. (2010)
and Leonidaki et al. (2013) for X-ray and optical SNRs. Also, it was target of the DDO80 search, which detected 2 SNRs. Matonick et al. (1997)
carried a relatively deep optical search for SNRs in this galaxy.
They detected 33 new SNRs, and confirmed two detections of DDO80. VLA observations (Turner \& Ho 1994, Eck, Cowan \& Branch 2002)
found only three radio SNR candidates. Pannuti, Schlegel \& Lacey (2007) searched for positional coincidences between
their sample of X-ray sources in NGC2403 and 35 SNRs of Matonick et al. (1997) and known radio SNRs in the galaxy. They found only two matches,
one with optical SNR, and one with the radio SNR candidate. In addition to that, Leonidaki et al. (2010) found 15 X-ray selected SNRs, while Leonidaki et al.
(2013) reported detection of 149 optical SNRs or SNR candidates. All optical SNRs from Matonick et al. (1997), except SNR-1, which was out of the field of view,
were detected by Leonidaki et al. (2013).

The total H$\alpha$ line flux density of 150 SNRs detected by Leonidaki et al. (2013) and Matonick et al. (1997) in NGC2403
is $5.65\times10^{-12}$ erg cm$^{-2}$ s$^{-1}$. This galaxy is the third by the number of optically detected SNRs, after M83 and M31,
but it is the first by the percentage of H$\alpha$ flux coming from the SNRs in comparison to total H$\alpha$ flux from the galaxy.

\subsection{M82}

M82 is a member of M81 galaxy group, with a distance of 3.53 Mpc.
It is one of the nearest and certainly the best-studied starburst
galaxy. The active starburst has continued for almost 20 Myr at a
rate of about 10 M$_{\sun} \rm{yr}^{-1}$. The evidence of such
strong star formation are also two (radio) SNe in the last 10
years. Recently, type Ia SN2014J was detected in this galaxy
(Fossey et al. 2014). Active starburst region is located in the
centre of the galaxy. This region was mostly avoided for optical
observations. Dust attenuation of light is restricting optical
detection of larger population of compact SNRs in the core of M82,
which have been studied at radio wavelengths (Kronberg \&
Wilkinson 1975; Kronberg, Biermann \& Schwab 1985; Huang et al.
1994; Muxlow et al. 1994; Fenech et al. 2008; Fenech et al. 2010).
Therefore, no SNRs have been detected at visible wavelengths in
M82 until de Grijs et al. (2000) have analyzed \textit{HST}
H$\alpha$ narrowband images. In part of the galaxy called M82B,
which is considered to be a "fossil" starburst region, they have
detected 10 SNR candidates. Interesting about optical SNR
detection from de Grijs et al. (2000) is that it was not based on
\hbox{[S\,{\sc ii}]}$/$H$\alpha$ ratio, but on the conditions that
H$\alpha$ luminosity of SNR should be less than  $14\times10^{35}$
erg s$^{-1}$ and that diameter of an SNR should be less than
100pc.

The total H$\alpha$ line flux density of 10 SNRs detected by de Grijs et al. (2000) is $2.20\times10^{-15}$ erg cm$^{-2}$ s$^{-1}$.

\subsection{M81}

M81 is a nearby spiral galaxy, the brightest galaxy in the Group.
M81 forms the most conspicuous physical pair with its neighbor,
M82, with whom it had a close encounter a few tens of million
years ago. The galaxy also contains a low luminosity AGN (Markoff
et al. 2008). The only thorough search for SNRs in M81 was done by
Matonick \& Fesen (1997). In five fields of view, covering the
whole disk of this galaxy, and by using the \hbox{[S\,{\sc
ii}]}$/$H$\alpha$ ratio technique, they detected 41 optical SNRs.
Ten years earlier, in a search for giant {\hbox{H\,{\sc ii}}}
regions in M81 using the VLA at 6 cm and 20 cm, Kaufman et al.
(1987) proposed that five sources could be SNRs, according to
their spectral indices. The latest paper on SNRs in M81 galaxy was
by Pannuti et al. (2007). They searched for X-ray counterparts to
optical and candidate radio SNRs using observations made with the
\textit{Chandra} X-Ray Observatory. Their field of view covered
only 23 optical and three radio candidate SNRs, but they found no
\textit{Chandra}-detected counterparts.

The total H$\alpha$ line flux density of 41 optical SNRs detected by Matonick \& Fesen (1997) is $1.8\times10^{-13}$ erg cm$^{-2}$ s$^{-1}$.

\subsection{NGC3077}

NGC3077 is a nearby dwarf galaxy, member of the M81 galaxy group.
The tidal interaction between galaxies in this group leads to
enhanced star formation (Walter et al. 2002). Concerning SNRs,
firstly there were only three X-ray candidates which were detected
using \textit{Chandra} observations (Ott, Martin \& Walter 2003).
Rosa-Gonzales (2005) found that one of these sources coincides
with one of radio sources they reported. Later, Leonidaki et al.
(2010) detected 5 SNRs candidates, using archival \textit{Chandra}
images, and among their five detections two were already known
(from Ott et al. 2003) and three were new. Finally, Leonidaki et
al. (2013) detected 24 optical SNRs in NGC3077, of which one had
both optical and X counterparts.

Here we also mention Garland, tidal arm, located eastward of the
galactic centre, where VLA HI observations (Walter et al. 2002)
showed that 90 \% of atomic hydrogen around NGC3077 is located.
Also, Karachentsev \& Kaisin (2007) reported that Garland has the
highest SFR per luminosity among 150 galaxies of the Local volume
with known SFRs. For those reasons, Andjeli\'{c} et al. (2011)
searched for optical SNRs in Garland, but they found no such
candidates.

The total H$\alpha$ line flux density of 24 optical SNRs detected by Leonidaki et al. (2013) in NGC3077 is $2.47\times10^{-13}$ erg cm$^{-2}$ s$^{-1}$.

\subsection{NGC7793}

NGC7793 is a member of nearby Sculptor Group, with a distance of
3.91 Mpc and an inclination angle of $\sim$50$^{\circ}$ (Tully
1988). Its spiral pattern is nearly lost in a general confusion of
{\hbox{H\,{\sc ii}}} and star formation regions. The SNR
population of NGC7793 has been well studied through optical, radio
and X surveys (Blair \& Long 1997, Pannuti et al. 2002, Pannuti et
al. 2011). According to these searches, there are 32 detected SNRs
(or SNR candidates) in total. First, Blair \& Long (1997) reported
detection of 28 SNRs based on optical search through narrowband
H$\alpha$ and \hbox{[S\,{\sc ii}]} filters. Afterwards, Pannuti et
al. (2002) detected five new radio SNR candidates, and two SNRs
from Blair \& Long (1997) list using their own VLA observations.
They also examined archival \textit{ROSAT} satellite data and
found no additional SNR candidates, but it should be kept in mind
that the search for SNRs in this range is complicated by the
presence of considerable diffuse X-ray emission throughout the
entire disk of this galaxy. Finally, Pannuti et al. (2011)
improved X-ray data for NGC7793 using \textit{Chandra}
observatory, but they only found one SNR previously detected in
radio, and one SNR - S11 (from Blair \& Long 1997) detected both
at optical and radio wavelengths. Another interesting object in
this galaxy is SNR candidate S26 from Blair \& Long 91997). {This
is now a known microquasar with surrounding nebula, studied by
several authors (Pakull et al. 2010, Soria et al. 2010, Dopita et
al. 2012).} It is an extended region with diameter of
approximately 260 pc, and with high \hbox{[S\,{\sc
ii}]}$/$H$\alpha$ ratio, prominent in H$\alpha$ line, but not a
tracer of star formation.

Blair \& Long (1997) pointed that NGC7793 is galaxy with not so
obvious a "gap" between \hbox{[S\,{\sc ii}]}$/$H$\alpha$ ratios
for {\hbox{H\,{\sc ii}}} regions and SNRs, which could lead to
numerous contaminations or misidentification of optical SNRs.

The total integrated H$\alpha$ line flux density from {26} SNRs
for Blair \& Long (1997) is ${2.76}\times10^{-13}$ erg cm$^{-2}$
s$^{-1}$.

\subsection{NGC4449}

NGC4449 is a Magellanic-type irregular galaxy with several areas
of recent star formation, including the nucleus and the bar. This
galaxy is known for an extensively studied Cas A-like SNR, which
has been observed through all wavelengths (optical: Balick \&
Heckman 1978; Blair, Kirshner \& Winkler 1983; X-rays: Patnaude \&
Fesen 2003; radio: Seaquist \& Bignell 1978; Lacey, Goss \&
Mizouni 2007). This SNR was the only optical SNR in NGC4449 until
recently, when Leonidaki et al. (2013) detected additional 69
optical candidate SNRs. In radio domain, Chomiuk \& Wilcots (2009)
contributed 7 new SNR detections, which were all also detected in
optical by Leonidaki et al. (2013). In X-rays, Summers et al.
(2003) observed NGC4449 using \textit{Chandra} observatory, and
they suggested two sources to be SNRs, while other 8 were
classified as SNR/XRB (X-ray binary). Using the same
\textit{Chandra} observations, Leonidaki et al. (2010) suggested
that 4 objects should be SNRs or SNR candidates. All of these 4
objects were also detected as optical SNR candidates (Leonidaki et
al. 2013).

The total H$\alpha$ line flux density of 70 optical SNRs detected by Leonidaki et al. (2013) in NGC4449 is $1.19\times10^{-12}$ erg cm$^{-2}$ s$^{-1}$.

\subsection{M83}

M83 is a grand-design spiral galaxy, with a starburst nucleus and
active star formation. Its almost face-on orientation gives us an
opportunity to have a detailed view of the processes of star
formation and destruction. It had six SNe in the last century,
which makes it second only to NGC6946. That is why we should not
be surprised that this galaxy is the one with the highest number
of optically detected SNR candidates. The first who searched for
optical SNRs in M83 were Blair \& Long (2004). They made
observations by using 4 m Blanco telescope at the Cerro Tololo
Inter-American Observatory and found 71 candidate SNRs. They also
searched for X-ray counterparts to their optical detections from
Soria \& Wu (2003) and found 15 position matches. Two years later,
Maddox et al. (2006) published VLA observations of M83, and they
found four sources that matched optical SNRs from Blair \& Long
(2004). Later on, Dopita et al. (2010), by using a Wide Field
Camera (WFC) on \textit{HST}, detected 60 SNRs in one field of
view. Of those 60 SNRs, one is a historical remnant, 40 are
detected in a classical way by using \hbox{[S\,{\sc
ii}]}$/$H$\alpha$ ratio in the disk region of M83, while other 19
were detected in ``nuclear" region ($R<$300 pc). SNRs in nucleus
were detected based on enhanced \hbox{[O\,{\sc ii}]} line, and
only 4 of them had measured H$\alpha$ fluxes. Among detections of
Dopita et al. (2010), there were 12 SNRs from Blair \& Long
(2004). The largest contribution to the number of observed optical
SNRs in M83 was made by Blair et al. (2012) {(see also erratum to
this paper
 by Blair et al. 2013)}. They observed the full extent of the galaxy, using Magellan I 6.5 m telescope under conditions of excellent
 seeing
and with extremely deep exposures of 70 minutes. They detected 225
ISM-dominated SNRs in M83, using \hbox{[S\,{\sc ii}]}$/$H$\alpha$
ratio technique, with additional 33 [OIII] selected objects, which
could be candidates for ejecta-dominated SNRs. They confirmed all
but three SNRs from Blair \& Long (2004), and 25 out of 40 disk
SNRs from Dopita et al. (2010). Blair et al. (2012) also searched
for X-ray counterparts to their optical SNRs, and found that 65 of
225 ISM-dominated SNRs have X-ray counterparts, as well as five
ejecta-dominated SNRs. {Just recently, Blair et al. (2014)
presented an expanded \emph{HST} survey of M83. They reported
detection of 26 new SNRs.} It should be mentioned that in all of
these three papers (Blair \& Long 2004, Dopita et al. 2010, Blair
et al. 2012, Blair et al. 2014) extensive analysis of physical and
statistical parameters of the SNRs sample in M83 is given.

The total H$\alpha$ flux from {296} SNRs in M83 which have
measured H$\alpha$ fluxes -- 225 ISM-dominated and 33
ejecta-dominated from Blair et al. 2012; 12 disk SNRs and 4
nuclear from Dopita et al. (2010), which are not detected by Blair
et al. (2012), {and 22 from Blair et al. (2014)}
 -- is ${6.39}\times10^{-12}$ erg cm$^{-2}$ s$^{-1}$.

\subsection{NGC4395}

NGC4395 is a nearby irregular starburst galaxy, hosting the least
luminous Seyfert nucleus. This galaxy may host one radio candidate
SNR, upon its non-thermal radio spectrum from VLA observations
(Sramek 1992, Vukoti\'{c} et al. 2005) and because of its
positional coincidence with {\hbox{H\,{\sc ii}}} region from Roy
et al. (1996). Unfortunately, this object was out of the
\textit{Chandra} field in which Leonidaki et al. (2010) suggested
two X-ray SNRs. At optical wavelengths, Leonidaki et al. (2013)
found 47 candidate SNRs, although their observations did not cover
the full extent of the galaxy. Among these detections there was
one X-ray candidate (from Leonidaki et al. 2010), while the other
was out of the field of view. The only radio candidate SNR in this
galaxy was also detected, but it had \hbox{[S\,{\sc
ii}]}$/$H$\alpha$ ratio less than 0.3.

The total H$\alpha$ line flux density of 47 optical SNRs detected by Leonidaki et al. (2013) in NGC4395 is $2.66\times10^{-13}$ erg cm$^{-2}$ s$^{-1}$.

\subsection{NGC5204}

NGC5204 is a nearby Magellanic-type galaxy, a member of the M101
group of galaxies. In the literature, most of the studies are
focused on the ULX source NGC5204 X-1, close to the centre of the
galaxy and its optical counterpart. In this galaxy, there are only
optical SNR detections. First, Matonick \& Fesen (1997) detected
three optical SNR candidates, while recently, Leonidaki et al.
(2013) confirmed these three detections and added 33 possible new
SNRs. In X-rays, Leonidaki et al. (2010) searched through archival
\textit{Chandra} images for SNRs in this galaxy, but they found no
such candidates.

The total H$\alpha$ line flux density of 36 optical SNRs detected by Leonidaki et al. (2013) in NGC5204 is $2.32\times10^{-13}$ erg cm$^{-2}$ s$^{-1}$.

\subsection{NGC5585}

NGC5585 is a late-type spiral galaxy in the M101 group. The only
paper which discussed SNRs in this galaxy is Matonick \& Fesen
(1997). They found only 5 optical SNRs. Among these 5 SNRs, there
is one interesting object, SNR 1, with extremely large dimensions
- $200\times90$ pc. This candidate SNR is similar to the already
mentioned microquasar S26 in NGC7793, although it is twice smaller
in size. Unfortunately, there are no additional observations of
this unusual object that could reveal its true nature.

The total integrated H$\alpha$ line flux density for 5 SNRs in NGC5585 from Matonick \& Fesen (1997) is
$6.78\times10^{-14}$ erg cm$^{-2}$ s$^{-1}$.

\subsection{NGC6946}

NGC6946 is a nearby (\textit{d}=5.9 Mpc) late type spiral galaxy
hiding a mild starburst nucleus (Turner \& Ho 1983). This galaxy
could be interesting for SNRs search because there are nine
historical SNe up to date in this galaxy (Barbon et al. 2010). The
first detected SNR in NGC6946 was MF16 (from Matonick \& Fesen
1997) and the detection was done by \textit{ROSAT} X-ray
observatory (Schlegel 1994) by following optical observations by
Blair \& Fesen (1994). The first systematic search for SNRs was
performed by Matonick \& Fesen (1997) in optical wavelengths. They
discovered 27 optical SNR candidates. The same year Lacey, Duric
\& Goss (1997) published their VLA observations of NGC6946, and
suggested 37 objects as possible SNRs. More recently, Pannuti et
al. (2007) have found, using their \textit{Chandra} observations,
six X-ray counterparts to candidate radio SNRs, and no
counterparts to the optical SNRs.

Due to its large dimensions ($>$100 pc) and its ultra high X-ray
luminosity, there are many papers debating true nature of MF16
source. Blair, Fesen \& Schlegel (2001), based on \textit{HST}
observations through narrowband filters, suggested that this
object could be a result of multiple SN explosions, close
temporally and spatially. Afterwards, Roberts \& Colbert (2003)
suggested upon \textit{Chandra} observations that extraordinary
X-ray luminosity of the MF16 arises from a black hole X-ray binary
and that this object is ULX source.

The total integrated H$\alpha$ line flux density from {26} SNRs in
NGC6946 from Matonick \& Fesen (1997) is ${1.28}\times10^{-13}$
erg cm$^{-2}$ s$^{-1}$.

\subsection{M101}

M101 is a nearly face-on ($\sim22^{\circ}$ inclination) giant
spiral galaxy at a distance of 6.7 Mpc. This galaxy is one of the
well studied for SNRs. The first major search for SNRs in M101 was
performed by Matonick \& Fesen (1997), when they reported
detection of 93 SNRs. With five fields of view, they covered the
full size of the galaxy. Before that, one candidate radio SNR in
M101, NGC5471B, a part of a giant {\hbox{H\,{\sc ii}}} region, was
known (Skillman 1985; Sramek \& Weedman 1986; Yang, Skillman \&
Sramek 1994). Afterwards, Pannuti et al. (2007) searched for X-ray
counterparts to optically identified SNRs from Matonick \& Fesen
(1997). Their observations covered 44 SNRs, and they found that
six of them are visible also in X-rays. The latest paper on SNRs
in M101 was the one by Franchetti et al. (2012) who used archival
\textit{HST} H$\alpha$ images for 55 of the 93 SNR candidates
identified by Matonick \& Fesen (1997) and conducted detailed
analysis of their physical structure and nature. Also, they used
deep \textit{Chandra} X-ray mosaic of M101 made by Kuntz \&
Snowden (2010) to search for X-ray emission of SNR candidates and
found that 21 of the 55 SNR candidates in their study have X-ray
counterparts. They also checked if the rest 38 SNRs from Matonick
\& Fesen (1997), which do not have \textit{HST} images, radiate in
X-rays, and they found that 11 of them have X-ray counterparts.
From \textit{HST} images, they found that ten objects, out of 55
SNRs from Matonick \& Fesen (1997), are actually superbubbles, and
another ten are OB/{\hbox{H\,{\sc ii}}} associations. These ten we
excluded from the list of SNRs in M101.

Finally, in order to estimate H$\alpha$ flux from SNRs in M101, we
combine the results from Matonick \& Fesen (1997) and Franchetti
et al. (2012). We have taken H$\alpha$ fluxes for 38 SNRs from
Matonick \& Fesen (1997), and for 35  SNRs Franchetti et al.
(2012). Total integrated H$\alpha$ line flux density for 73 SNRs
and in M101 is ${3.46}\times10^{-13}$ erg cm$^{-2}$ s$^{-1}$.

\subsection{M74}

M74 (NGC628) is a face-on spiral galaxy at a distance of 7.3 Mpc
and is the brightest member of a small M74 group of galaxies. This
galaxy has been well studied for its star-forming properties
(Leievre \& Roy 2000, Elmegreen et al. 2006). M74 also had three
SN explosions in the last twelve years. SNRs at optical
wavelengths in this galaxy were detected by Sonbas et al. (2010).
They detected nine SNRs in this galaxy. Previously, M74 was
observed in X-rays using the \textit{Chandra} and
\textit{XMM-Newton} data by Soria, Pian \& Mazzali (2004). Sonbas
et al. (2010) found three matches between their SNR candidates and
X-ray sources from Soria et al. (2004).

The total integrated H$\alpha$ line flux density for 9 SNRs in M74 from Sonbas et al. (2010) is
$1.05\times10^{-13}$ erg cm$^{-2}$ s$^{-1}$.

\subsection{NGC2903}

NGC2903 is a late-type barred spiral galaxy at a distance of 8.9 Mpc.
It is also known as a ``hot-spot" galaxy, because of its numerous bright nuclear condensations (Sersic 1973)
which are believed to be OB associations and clusters. The only SNRs search in this galaxy was performed
by Sonbas, Akuyz \& Balman (2009). They detected five optical SNR candidates. Besides that, Tsai et al. (2006)
reported in their subarcsecond-resolution VLA imaging of NGC2903 that one of seven detected discrete
radio sources could be an SNR.

The total integrated H$\alpha$ line flux density for 5 SNRs in NGC2903 from Sonbas et al. (2009) is
$5.88\times10^{-14}$ erg cm$^{-2}$ s$^{-1}$.

\section{Discussion}

In previous section we gave an overview on the achievements in
optical detection of SNRs in nearby galaxies, as well as some
notice on detection of SNRs at other wavelengths. {In our sample,
we present 18 galaxies which have been surveyed for optical SNRs.}

In Table 2 we give the number and total H$\alpha$ flux of optical
SNRs in each galaxy from our sample. Also, we present data taken
from Kennicutt et al. (2008) which give total H$\alpha$ flux of
the galaxy, which can be used to estimate SFR (according to
equation (1)). Kennicutt et al. (2008) have published ``An
H$\alpha$ imaging survey of galaxies in the local 11 Mpc volume",
which provided us with the H$\alpha$ fluxes for {18} galaxies in
our sample. {Fluxes from Kennicutt et al. (2008) are corrected for
Galactic foreground extinction and \hbox{[N\,{\sc ii}]}
contamination. In Table 2 we give values used in Kennicutt et al.
(2008) for these corrections. We applied these same corrections
for SNRs' fluxes taken from literature. In the cases where
H$\alpha$ fluxes for SNRs given in original work have been
previously corrected for Galactic foreground extinction and
\hbox{[N\,{\sc ii}]} contamination, we recalculate fluxes in order
to make them corrected in the same manner as was done by Kennicutt
et al. (2008). In order to remove \hbox{[N\,{\sc ii}]}
contamination from fluxes we used integrated \hbox{[N\,{\sc ii}]}
$\lambda\lambda$6548, 6583/H$\alpha$ ratio from Kennicutt et al.
(2008) and the procedure presented in Vu\v{c}eti\'{c} et al.
(2013), which takes into account different filter transmittance at
position of each of these three emission lines. No internal
extinction correction was done to any of the fluxes presented in
this paper.}

\setcounter{table}{1}
\begin{table*}
 \caption{Our sample of galaxies with optical SNRs. The entries are organized as follows.
Column (1): Galaxy name, as in Table 1. Column (2): Number of
optically detected SNRs in the galaxy,{ with determined H$\alpha$
fluxes.} {Column (3): Fraction of galactic angular extension which
has been surveyed for optical SNRs.} Column (4): H$\alpha$ flux of
optically detected SNRs, {corrected for Galactic foreground
extinction and \hbox{[N\,{\sc ii}]} contamination as it was done
in Kennicutt et al. (2008).} {Column (5): Fractional error of
H$\alpha$ flux of optically detected SNRs, as reported in
literature. Slash sign means that no error estimates were
published.} {Column (6): H$\alpha$ luminosity of optically
detected SNRs, corrected for Galactic foreground extinction and
\hbox{[N\,{\sc ii}]} contamination same as in Kennicutt et al.
(2008).} Column (7): The adopted integrated \hbox{[N\,{\sc ii}]}
$\lambda\lambda$6548, 6583/H$\alpha$ ratio (sum of both
components), used in Kennicutt et al. (2008). Column (8): B-band
Galactic foreground extinction, used in Kennicutt et al. (2008).
Extinction in H$\alpha$ band is calculated as
$A_{\rm{H}\alpha}=0.6A_{B}$. Column (9): Integrated H$\alpha$ flux
for the galaxy. The flux is calculated using integrated H$\alpha$
luminosity and distance listed in Tables 2 and 3 in Kennicutt et
al. (2008). {This flux is reported with median fractional error of
12\%.} {Column (10): Integrated H$\alpha$ luminosity for the
galaxy.} Column (11): Percentage of H$\alpha$ flux coming from the
SNRs in comparison to total H$\alpha$ flux of the galaxy -- R.
Since SFR is directly proportional to the H$\alpha$ luminosity
(and flux) (see Equation 1), this percentage also represents
fractional error which is made when H$\alpha$ flux coming from
SNRs is not removed from total H$\alpha$ flux of the galaxy.
{Column (12): Fractional error of R. It represents the sum of
fractional errors for H$\alpha$ fluxes of SNRs and of the galaxy.}
Column (13): references from which H$\alpha$ fluxes of optically
detected SNRs are taken. }
 \label{symbols}
 \begin{tabular}{@{\extracolsep{-2.5mm}}lccccccccccccc}
  \hline
Galaxy  &   Num. of &   Fraction & F$_{\rm{SNRs}}$  &   $\delta$F$_{\rm{SNRs}}$ &   L$_{\rm{SNRs}}$ &
\hbox{[N\,{\sc ii}]}/H$\alpha$ $^{*}$ &    $A_{{B}}$ $^{*}$   &  F$_{\rm{gal}}$ $^{*}$ & L$_{\rm{gal}}$ $^{*}$ & R    & $\delta$R&    Ref. \\
name    & optical   & surveyed  & (erg cm$^{-2}$ s$^{-1}$)& (\%)    & (erg s$^{-1}$)    &   ratio   &   (mag)   & (erg cm$^{-2}$ s$^{-1}$)  &
(erg s$^{-1}$) &   (\%) & (\%) \\
&SNRs & & $\times10^{-14}$ & &$\times10^{38}$ & & &$\times10^{-12}$ & $\times10^{39}$&  & & \\
(1) &   (2) &   (3) &   (4) &   (5) &   (6) &   (7) &   (8) &   (9) &   (10) & (11) & (12) & (13) \\
\hline
M31 &   156 &   1   &   371.9   &   /   &   2.8 &   0.54    &   0.18    &   360.4   &   26.9    &   1.0    &   12  &   1   \\
M33 &   137 &   1   &   544.6   &   /   &   4.6 &   0.27    &   0.18    &   383.0   &   32.4    &   1.4    &   12  &   2   \\
NGC300  &   22  &   1   &   23.1    &   22  &   1.1 &   0.2 &   0.06    &   31.6    &   15.1    &   0.7    &   34  &   3   \\
NGC4214 &   92  &   1   &   178.0   &   2   &   18.2    &   0.16    &   0.05    &   15.2    &   15.5    &   11.7   &   14  &   4   \\
NGC2403 &   150 &   0.88    &   620.7   &   1   &   77.0    &   0.29    &   0.17    &   48.6    &   60.3    &   12.8   &   13  &   4,5 \\
M82 &   10  &   0.07    &   2.7 &   5   &   0.4 &   0.3 &   0.4 &   78.8    &   117.5   &   0.1    &   17  &   6   \\
M81 &   41  &   1   &   18.8    &   /   &   3.0 &   0.51    &   0.24    &   37.3    &   58.9    &   0.5    &   12  &   7   \\
NGC3077 &   24  &   1   &   28.3    &   6   &   5.0 &   0.38    &   0.25    &   5.5 &   9.5 &   5.2    &   18  &   4   \\
NGC7793 &   27  &   1   &   28.8    &   /   &   5.3 &   0.25    &   0.08    &   20.8    &   38.0    &   1.4    &   12  &   8   \\
NGC4449 &   71  &   1   &   121.6   &   2   &   25.8    &   0.23    &   0.04    &   24.2    &   51.3    &   5.0    &   14  &   4   \\
M83 &   296 &   1   &   653.0   &   /   &   156.0   &   0.53    &   0.21    &   74.4    &   177.8   &   8.8    &   12  &   9,10,11 \\
NGC4395 &   47  &   0.73    &   27.2    &   3   &   6.9 &   0.19    &   0.04    &   4.5 &   11.5    &   6.0    &   15  &   4   \\
NGC5204 &   36  &   1   &   23.6    &   4   &   6.1 &   0.13    &   0.03    &   3.1 &   8.1 &   7.5   &   16  &   4   \\
NGC5585 &   5   &   1   &   6.9 &   /   &   2.7 &   0.18    &   0.03    &   2.1 &   8.1 &   3.3    &   12  &   7   \\
NGC6946 &   26  &   0.95    &   12.1    &   /   &   5.0 &   0.54    &   1.54    &   69.2    &   288.4   &   0.2    &   12  &   7   \\
M101    &   73  &   0.98    &   35.0    &   /   &   19.2    &   0.54    &   0.02    &   39.8    &   213.8   &   0.8    &   12  &   7,12    \\
M74 &   9   &   0.83    &   9.4 &   /   &   6.0 &   0.4 &   0.21    &   11.6    &   74.1    &   0.8    &   12  &   13  \\
NGC2903 &   5   &   1   &   4.9 &   /   &   4.6 &   0.56    &   0.1 &   12.4    &   117.5   &   0.4    &   12  &   14  \\
\hline

 \end{tabular}
\begin{flushleft}
$^{*}$ From Kennicutt et al. (2008).\\
\textsc{References:} (1) Lee \& Lee 2014; (2) Long et al. 2010; (3) Millar et al. 2011; (4) Leonidaki et al. 2013; (5) Matonick et al. 1997;
(6) de Grijs et al. 2000; (7) Matonick  \& Fesen 1997; (8)  Blair \& Long 1997; (9) Dopita et al. 2010; (10) Blair et al. 2013; (11) Blair et al. 2014
(12) Franchetti et al. 2012; (13) Sonbas et al. 2010; (14) Sonbas et al. 2009.
\end{flushleft}

 \end{table*}

 Table 2 shows us that the percentage of H$\alpha$ emission originating from the SNR contamination
is within the range of 0.1--13\%. From Column (11) of Table 2 we
see that six out of seven highest values in this column originate
from the paper by Leonidaki et al. (2013), and are within the
range of 5--13\%. Such a high percentage of around 10\% is not a
negligible source of error in determined SFRs. The situation that
six out of seven highest values are coming from the same group of
authors does introduce doubt in their plausibility. It could be
that the source of this high efficiency in the detection of SNRs
in Leonidaki et al. (2013) is because they used a somewhat milder
\hbox{[S\,{\sc ii}]}$/$H$\alpha$$>0.3$ criterion for SNR
detection. This lower criterion was also used in Dopita et al.
(2010). Also, we add to this that Leonidaki et al. (2013) detected
150 SNRs in NGC2403, while Matonick et al. (1997) detected only
35. When we compare differences in telescopes and exposure times
used by these two groups, it is expected to have two times higher
detection efficiency by Matonick et al. (1997) than by Leonidaki
et al. (2013). Although, one should have in mind that Matonick et
al. (1997) used a stronger \hbox{[S\,{\sc ii}]}$/$H$\alpha$$>0.45$
criterion.

M83 galaxy is the {third} on the list of the highest percentage of
SNR H$\alpha$ emission contamination -- {9}\%. This galaxy boasts
the largest number of optical SNRs ({296}), and is the best
sampled for optical SNRs among all galaxies. It is not something
to be surprised at, when one knows that M83 was observed with 4 m
Blanco telescope (Blair \& Long 2004), then with \textit{HST}
(Dopita et al. 2010, {Blair et al. 2014}) and with Magellan I 6.5m
telescope with very long exposures (Blair et al. 2012). For this
reason, we think that the percentage similar to the one found in
M83 is expected to be somewhat closer to the real contribution of
the SNRs emission to the total H$\alpha$ emission in spiral
galaxies.

We emphasize that in our estimates of the H$\alpha$ fluxes from
SNRs in each galaxy, we have taken fluxes from both confirmed SNRs
(confirmed either spectroscopically or by observations through
other wavelengths) and SNR candidates, which have been classified
only upon the \hbox{[S\,{\sc ii}]}$/$H$\alpha$ ratio criterion.
{This could lead to the overestimation of percentage R,
considering that there might be numerous false identifications of
SNR candidates.}

{In Column (3) of Table 2 we present what fraction of angular
extension of a galaxy has been covered with observations for
optical SNRs. The majority of galaxies have been fully covered,
four of them have been more than 80\% covered, while M82 galaxy
has only a small fraction of disk observed for SNRs. In galaxies
which have not been fully observed, only the outer parts of
galactic surface have not been covered. If we assume that SNe
follow radial distribution of gas and dust in galaxy (Hatano,
Branch \& Deaton 1998), then we expect that the number of SNe, as
well as the number of SNRs, will rapidly decrease as we move away
from the galactic center. In this context, as only the outer parts
of galaxies have not been observed, we do not expect a major
change of number of SNRs and their total H$\alpha$ flux if the
whole galactic surface were to be observed. Only in the case of
M82 the change would be significant, but due to a large internal
extinction in that galaxy, it is hard to estimate the number of
SNRs that would be detected in the galaxy as a whole.}

\begin{figure*}
\centerline{\includegraphics[bb= -113 -191 722
984,width=0.925\textwidth,keepaspectratio]{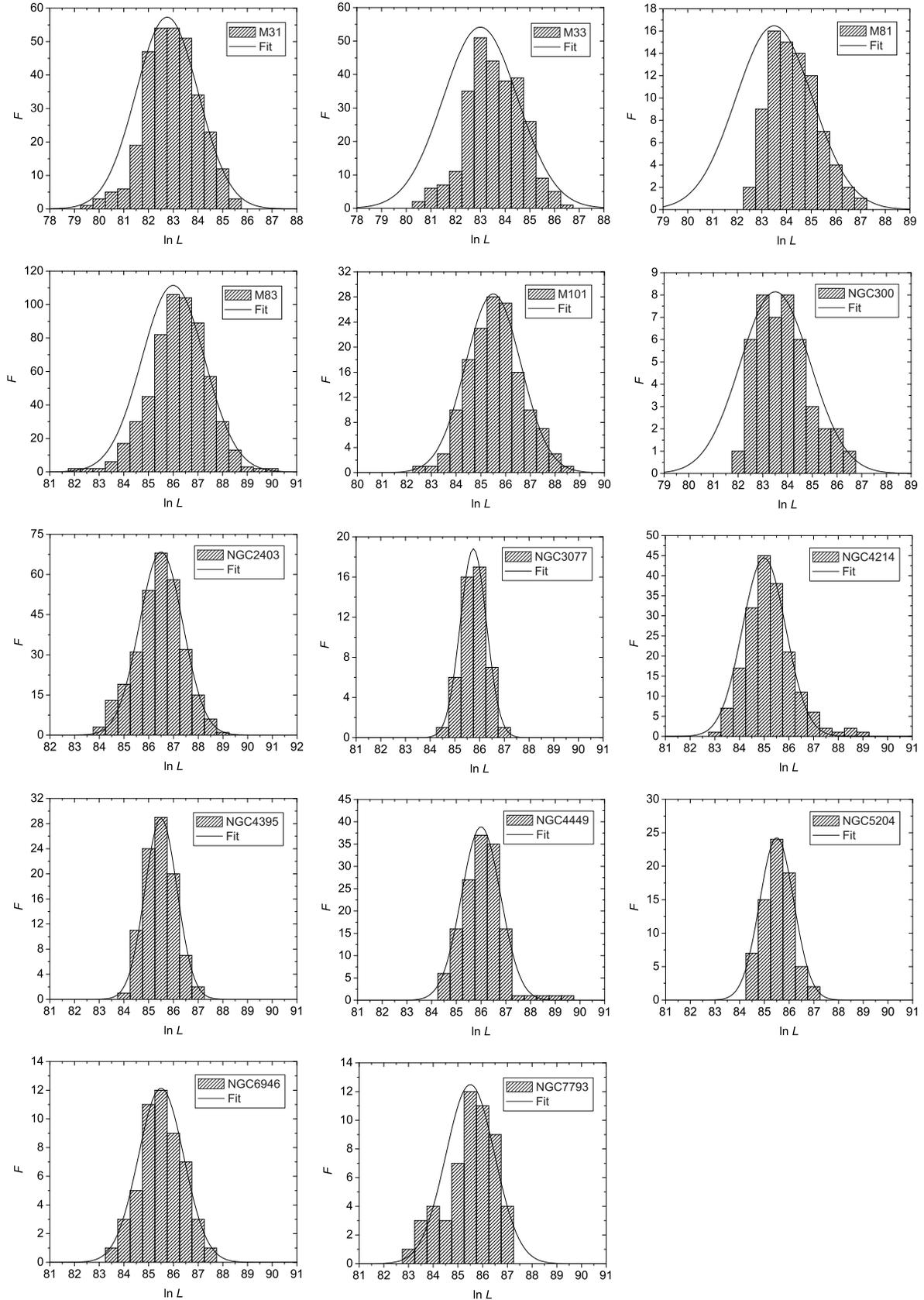}}
\begin{center}
\caption{Number distributions of SNRs in our galaxy sample and fitted normal distributions.}
\end{center}
\end{figure*}

{Another very important question which we discuss here is how  we
can compare data collected in very different studies, with
different flux sensitivities and from different galaxies.}

{In order to discuss the completeness of the samples, we have to
assume some number distribution function of SNRs with respect to
flux or luminosity -- {we shall choose luminosity so than we can
compare different samples.} If this distribution is log-normal
i.e. normal or gaussian in the logarithmic scale, $F(L) =
\frac{N^*}{\sigma \sqrt{2 \pi}} e^{-(\ln L - \mu)^2/(2\sigma
^2)}$, then from the fit we can estimate the number of missing
SNRs $N^* - N$, where $N^*$ and $N$ are the corrected and detected
number of SNRs, and their contribution to the total H$\alpha$
luminosity. $\mu = \ln L_{\mathrm{m}}$ is the natural logarithm of
luminosity which corresponds to the maximum of the distribution
i.e. the mean and $\sigma$ is the standard deviation of normal
distribution.}

{It is usually assumed that a sample is complete for high fluxes
(or luminosities) up to the maximum (Leonidaki et al. 2013;
Matonick \& Fesen 1997). While it is quite likely that we will
detect all bright and luminous objects, we can still miss the
extended low surface-brightness (and thereby luminous) SNRs.
However, we will assume that these are rare.}

{We have thus fitted only the right-hand side of the observed
distributions, excluding local minima and bars which are lower
than their left-hand side counterparts and keeping the mean i.e.
$\ln L_{\mathrm{m}}$ fixed. We could do this only for 14 galaxies
from our sample, which have enough SNRs detected in order to have
the
 distribution of the number of SNRs over luminosities.
The results are given in Fig 1 and
Table 3. When fit parameters $N^*$ and $\sigma$ have been found we
can obtain the corrected total luminosity of SNRs which is
actually the mean of the log-normal distribution
\begin{equation}
L^* = \int _0 ^\infty L f(L) dL = N^* \cdot e^{\mu +\sigma ^2/2} ,
\end{equation}
where $f(L)= \frac{N^*}{\sigma \sqrt{2 \pi} L} e^{-(\ln L -
\mu)^2/(2\sigma ^2)}$.}
\vspace{2mm}

{Now we can also obtain ratio of H$\alpha$ luminosity coming from
the SNRs to the total H$\alpha$ luminosity of the galaxy, which is
corrected for possible incompleteness of the SNR sample -- R$^*$.
If we compare uncorrected R and corrected R$^*$ (from Table 3) we
can see that there is the biggest change in values for M33 and
NGC4214 galaxies, but highest value of 13\% of SNR contamination
is not changed.}

\begin{table*}
 \centering
 \begin{minipage}{168mm}
  \caption{Basic properties and corrected parameters of number distributions of SNRs in our galaxy sample. $N$ is detected number of SNRs in each galaxy,
   $N^*$ and $\Delta N^*$ are the corrected number and its error from Gaussian fit, $L_{\mathrm{o}}$ is the minimum luminosity observed and $L_{\mathrm{m}}$
   is the luminosity which corresponds to the maximum of the distribution, $\sigma$    and $\Delta \sigma$ are standard deviation of normal distribution and
   its error, $L_{\mathrm{SNRs}}$ is measured total luminosity of SNRs, $L^*$ the corrected total luminosity of SNRs, R and R$^*$ are ratio and the corrected
    ratio of H$\alpha$ luminosity coming from the SNRs to the total H$\alpha$ luminosity of the galaxy.}
 \vspace{5mm}
  \center{
  \begin{tabular}{@{\extracolsep{2.0mm}}lccccccccccc@{}}
 \hline
Galaxy  &   $N$ &   $N^*$   &   $\Delta N^*$    &   ln
$L_{\mathrm{o}}$ &   ln $L_{\mathrm{m}}$ &   $\sigma$    & $\Delta
\sigma$ &   $L_{\mathrm{SNRs}}$ (erg s$^{-1}$)  &
  $L^*$  (erg s$^{-1}$)  & R  &  R$^*$ \\
\hline
M31 &   156 &   181 &   6   &   79.51   &   82.75   &   1.26    &   0.05    &   2.76$\times 10^{38}$    &   3.47$\times 10^{38}$    &   1.03 &   1.29    \\
M33 &   137 &   205 &   15  &   80.53   &   83  &   1.51    &   0.11    &   4.60$\times 10^{38}$    &   7.13$\times 10^{38}$        & 1.42 & 2.20      \\
M81 &   41  &   64  &   3   &   82.89   &   83.5    &   1.55    &   0.07    &   2.99$\times 10^{38}$    &   3.90$\times 10^{38}$       & 0.50 &  0.66     \\
M83 &   296 &   355 &   10  &   82.12   &   86  &   1.27    &   0.07    &   1.56$\times 10^{40}$    &   1.78$\times 10^{40}$        &  8.78 &  10.00  \\
M101    &   73  &   80  &   4   &   82.61   &   85.5    &   1.12    &   0.05    &   1.91$\times 10^{39}$    &   2.03$\times 10^{39}$     &  0.88 &  0.95     \\
NGC300  &   22  &   29  &   2   &   82.43   &   83.5    &   1.42    &   0.09    &   1.11$\times 10^{38}$    &   1.46$\times 10^{38}$      & 0.73 &    0.96   \\
NGC2403 &   150 &   151 &   3   &   84.06   &   86.5    &   0.88    &   0.03    &   7.71$\times 10^{39}$    &   8.20$\times 10^{39}$      & 12.77 &  13.60     \\
NGC3077 &   24  &   25  &   1   &   84.87   &   85.75   &   0.53    &   0.01    &   4.96$\times 10^{38}$    &   5.01$\times 10^{38}$       & 5.18 &    5.24   \\
NGC4214 &   92  &   98  &   4   &   83.28   &   85  &   0.88    &   0.04    &   1.82$\times 10^{39}$    &   1.19$\times 10^{39}$        &  11.71 & 7.66   \\
NGC4395 &   47  &   47  &   2   &   84.25   &   85.5    &   0.65    &   0.02    &   6.88$\times 10^{38}$    &   7.87$\times 10^{38}$        & 6.02 &  6.86    \\
NGC4449 &   71  &   77  &   6   &   84.73   &   86  &   0.79    &   0.08    &   2.57$\times 10^{39}$    &   2.35$\times 10^{39}$        &  5.03   & 4.58 \\
NGC5204 &   36  &   42  &   2   &   84.56   &   85.5    &   0.69    &   0.03    &   6.12$\times 10^{38}$    &   7.22$\times 10^{38}$        & 7.50 &   8.89  \\
NGC6946 &   26  &   28  &   1   &   83.63   &   85.5    &   0.92    &   0.03    &   5.05$\times 10^{38}$    &   5.80$\times 10^{38}$        &  0.17 &   0.20 \\
NGC7793 &   27  &   31  &   4   &   83.42   &   85.5    &   0.99    &   0.11    &   5.30$\times 10^{38}$    &   6.86$\times 10^{38}$        & 1.39   & 1.80  \\
\hline
\end{tabular}}
\end{minipage}
\end{table*}

{We can see that, according to the fit, only in galaxies M33, M81
and NGC300 there is a significant number of low luminous SNRs
missing. However, because of the characteristics of log-normal
distribution, $L^*$ is not very different from uncorrected total
H$\alpha$ luminosity of SNRs, neither for these three nor for all
other galaxies. For two galaxies, NGC4214 and NGC4449, $L^*$ is
slightly smaller than $L_{\mathrm{SNRs}}$, which is due to the
existence of very bright SNRs in these galaxies i.e. the
distribution can not be approximated with a single gaussian.}

{Even if the real distribution over luminosities is indeed
gaussian (in the logarithmic scale), which may be an overly
simplistic assumption, because of the selection effects and
sensitivity limits, the observed mean i.e. the logarithm of
luminosity which corresponds to the maximum of the distribution
$\ln L_{\mathrm{m}}$ may not be the true mean $\mu$, which in
reality may be shifted to lower luminosities. In principle, we may
expect that if the sensitivity limit i.e. the minimum luminosity
$L_{\mathrm{o}}$ is far from $L_{\mathrm{m}}$, the mode i.e. mean
is well determined. To test this, we have looked for a possible
correlation between the corrected ratio of H$\alpha$ luminosity
coming from the SNRs to the total H$\alpha$ luminosity of the
galaxy R$^*$ and $\ln ( L_{\mathrm{m}}/L_{\mathrm{o}})$ in spirals
-- galaxies with high $\ln ( L_{\mathrm{m}}/L_{\mathrm{o}})$
should then have a more reliable (and higher) R. However, no
evident correlation can be seen from Fig. 2. We have also checked
if there is any correlation between R$^*$ and galaxy properties -
inclination and galaxy type, but again without much success (Figs.
3 and 4).}

\begin{figure}
\centerline{\includegraphics[bb= 0 0 273 234,width=0.525\textwidth,keepaspectratio]{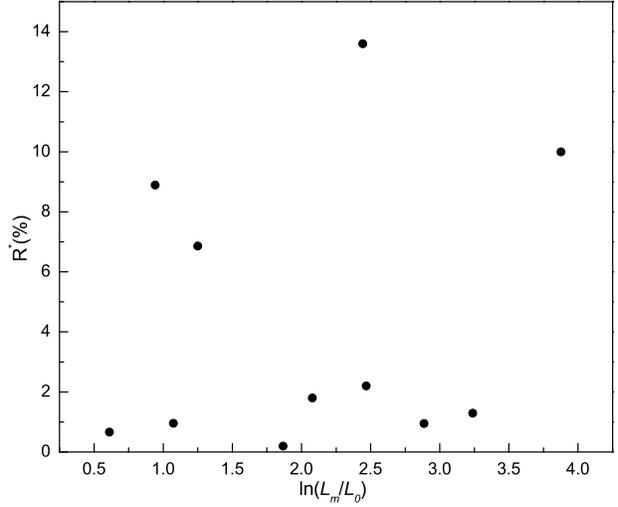}}
\begin{center}
\caption{Plot of corrected ratio of H$\alpha$ luminosity coming
from the SNRs to the total H$\alpha$ luminosity of the galaxy,
R$^*$, versus $\ln ( L_{\mathrm{m}}/L_{\mathrm{o}})$ for 11 spiral
galaxies from our sample. $L_{\mathrm{m}}$ is the luminosity that
corresponds to the maximum of SNRs number distribution and
$L_{\mathrm{o}}$ is the sensitivity limit i.e. the minimum
luminosity. No correlation is present.}
\end{center}
\end{figure}

\begin{figure}
\centerline{\includegraphics[bb= 0 0 271 232,width=0.525\textwidth,keepaspectratio]{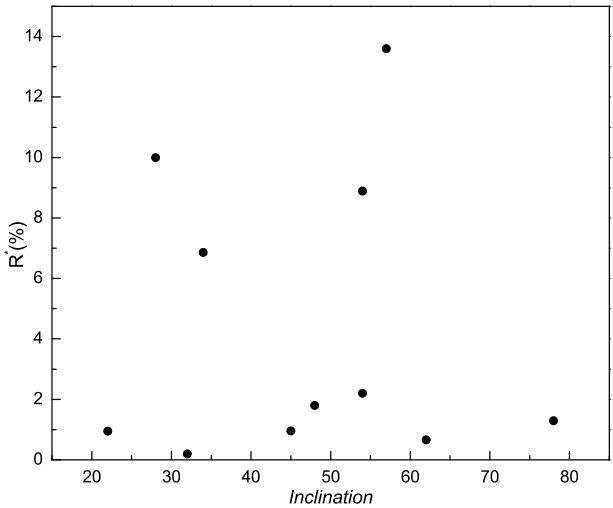}}
\begin{center}
\caption{Plot of corrected ratio of H$\alpha$ luminosity coming
from the SNRs to the total H$\alpha$ luminosity of the galaxy,
R$^*$, versus galaxy inclination for 11 spiral galaxies from our
sample. No correlation is present.}
\end{center}
\end{figure}

\begin{figure}
\centerline{\includegraphics[bb= 0 0 276 232,width=0.525\textwidth,keepaspectratio]{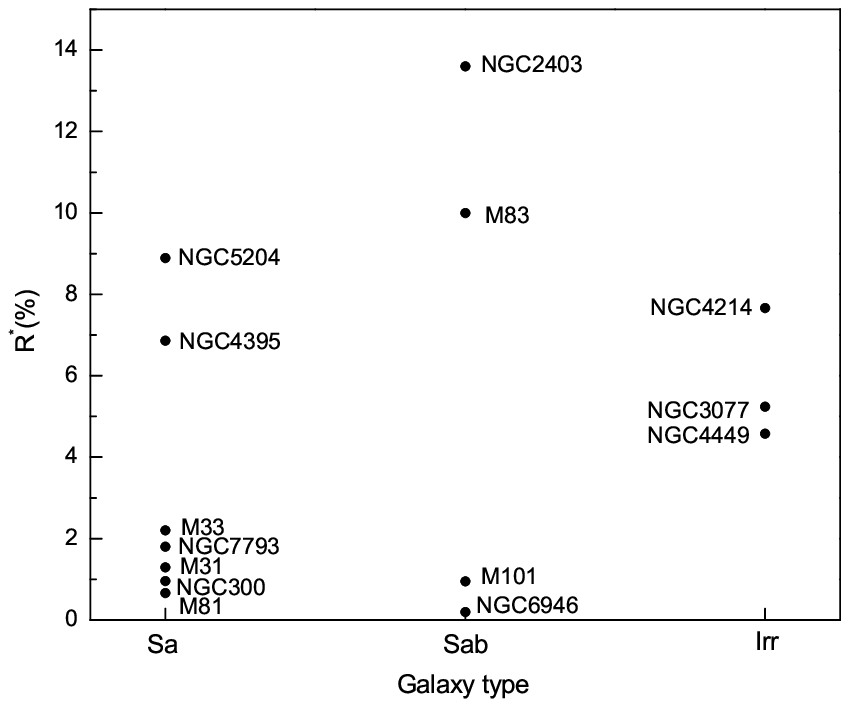}}
\begin{center}
\caption{Dependance of corrected ratio of H$\alpha$ luminosity
coming from the SNRs to the total H$\alpha$ luminosity of the
galaxy, R$^*$, on galaxy type.}
\end{center}
\end{figure}

\subsection{SNR contamination and SN type}
If we discuss different types of SN explosions and the remnants
they leave, we assume that remnants of type Ia SNe are not as
easily detectable as those of type II or Ib/c. Type Ia SNe have a
possible tendency to occur in the low-density regions of a galaxy
and leave no optically detectable remnants. Similarly,
Balmer-dominated SNRs which are thought to be related to type Ia
SNe (van den Bergh 1988) will be missed by optical searches using
\hbox{[S\,{\sc ii}]}/H$\alpha$$>$0.4 criterion.{ Study on SN rate
by Mannucci et al. (2005) found that there is about 25\% of type
Ia explosions in spiral and irregular galaxies. This is in good
agreement with Lee \& Lee (2014), who suppose that 23\% of all
optical SNRs in M31 originated from type Ia events.} Furthermore,
Braun \& Walterbos (1993) estimated that about half of all
core-collapse SNe (type II or Ib/c), will happen in associations
and leave no detectable remnant. {With these assumptions, we would
be able to optically detect only about one third of all SNRs.}
This occurs since only a few SNRs are observed inside
{\hbox{H\,{\sc ii}}} regions and because optical search appears
biased against detecting large and faint SNRs. Although SNRs are
rarely detected inside {\hbox{H\,{\sc ii}}} regions and stellar
associations, there must certainly be SNRs inside such objects
(remnants of type II and Ib/c SNe). There is evidence that SNRs
are often associated with giant {\hbox{H\,{\sc ii}}} regions -- in
30 Doradus giant {\hbox{H\,{\sc ii}}} region in LMC (Ye 1988), in
SMC (Ye, Turtle \& Kennicutt 1991), in NGC4449 (Kirshner \& Blair
1980). Even though SNRs embedded in an {\hbox{H\,{\sc ii}}} region
are not separated from the associated region, we expect that there
must be some contribution from such SNRs to the total H$\alpha$
radiation, {which should increase influence of SNR contamination
on SFRs.}

The selection effects that affect detection of SNRs are related to an observational fact that the optical SNRs on average lie in regions of lower gas density.
Due to this they are biased against star forming regions, which typically tend to have higher than  average gas density (for details on this observational bias
see Pannuti et al. 2000).
Taking this into account, we emphasize that derived percentages of the H$\alpha$ emission from the SNRs to the total H$\alpha$ fluxes used to estimate SFRs in
the galaxies represent only a lower limit.

\section{Conclusions}
{Out of 25 nearby galaxies with observed SNRs in optical range, we
have presented details on SNRs detection for our sample of 18
galaxies.} We have discussed the contribution of the H$\alpha$
fluxes from the SNRs to the total H$\alpha$ flux and its influence
to the derived SFR for each galaxy in the sample. We have found
that the {average SNR contamination to the total H$\alpha$
flux and derived SFRs for analysed nearby galaxies is  $5\pm 5$\%.
The highest SNR contamination is about 13\%.} M83 galaxy is best
sampled for optical SNRs and it has {9\%} of SNRs in total
H$\alpha$ emission. We expect that percentages similar to this one
should be close to the real contribution of the SNRs emission to
the total H$\alpha$ emission in spiral galaxies. Due to the
selection effects, the SNR H$\alpha$ contamination obtained in
this paper represent only a lower limit.

\section*{Acknowledgments}
{ We are grateful to Miroslav Filipovi\'{c} for useful comments
and suggestions, Dragana Momi\'{c} for careful reading and
correction of the manuscript and to the anonymous referee for a
comprehensive report that helped us improve it significantly.}
This research has been supported by the Ministry of Education,
Science and Technological Development of the Republic of Serbia
through project No. 176005 "Emission nebulae: structure and
evolution" and it is a part of the joint project of Serbian
Academy of Sciences and Arts and Bulgarian Academy of Sciences
"Optical search for supernova remnants and {\hbox{H\,{\sc ii}}}
regions in nearby galaxies (M81 group and IC342)". This research
has made use of the NASA/IPAC Extragalactic Database (NED) which
is operated by the Jet Propulsion Laboratory, California Institute
of Technology, under contract with the National Aeronautics and
Space Administration, and NASA's Astrophysics Data System
Bibliographic Services.

\bsp

\label{lastpage}

\appendix
\section{Optical SNRs in nearby galaxies}

This is a sample of Table A1. The complete Table A1, as well as the Tables A2-A18, are
available in Online Appendix.

\begin{table*}
 \centering
 \begin{minipage}{110mm}
 \caption{Optical SNRs in M31 galaxy$^{*}$. }
 \vspace{-1mm} {

 \begin{tabular}{@{\extracolsep{0mm}}lccccc@{}}
  \hline
  Object    &   RA  &   Dec.    &   F(H$\alpha$)    &   Diameter    &   {[S\,{\sc ii}]}/H$\alpha$  \\
name    &    (J2000.0)  &   (J2000.0)   &   (erg cm$^{-2}$ s$^{-1}$)    &   (\arcsec)   &   ratio   \\
& (\degr) & (\degr) & $\times10^{-15}$ & & \\
\hline
1   &   9.4056797   &   39.862778   &   7.8 &   12.9    &   0.86    \\
2   &   9.8472862   &   40.738834   &   51.5    &   15.3    &   1.06    \\
3   &   9.8783941   &   40.357998   &   7.1 &   8.8 &   0.96    \\
4   &   9.9373655   &   40.498367   &   31.8    &   12.6    &   1.09    \\
5   &   9.9593277   &   40.349838   &   43.9    &   14.8    &   0.40    \\
6   &   9.968482    &   40.495148   &   112.7   &   19.6    &   0.67    \\
... &   ...  &   ...   &   ...    &   ...    &   ...    \\
156 &   11.6789007  &   42.21674    &   9.2 &   13.6    &   1.12    \\
\hline
\end{tabular}}
\begin{flushleft} \vspace{-2mm}
$^{*}$Note: data taken from Lee \& Lee (2014).
\end{flushleft}
\end{minipage}
\end{table*}

\end{document}